\documentclass[pra,twocolumn,superscriptaddress,floatfix]{revtex4-1}
\usepackage{amsmath,amsfonts,amssymb,amsthm,graphics,graphicx,epsfig,bbm}
\usepackage[colorlinks=true,citecolor=blue,linkcolor=blue,urlcolor=blue]{hyperref}
\usepackage[usenames]{color}
\usepackage{graphicx}
\usepackage{subfigure}
\usepackage{amsmath}
\usepackage{epsfig}
\usepackage{dcolumn}
\usepackage{bm}
\usepackage{color}
\usepackage{verbatim}
\usepackage{epstopdf}
\usepackage{amssymb} 
\usepackage{amstext}
\usepackage{latexsym}
\usepackage{hyperref}
\usepackage{amsfonts}
\usepackage{psfrag}
\usepackage{xcolor}
\usepackage{times}
\usepackage{dsfont }

\newcommand{\ket}[1]{\ensuremath{\left|{#1}\right\rangle}}

\begin{document}
\title{Spectral collapse via two-phonon interactions in trapped ions}

\author{S. Felicetti}
\affiliation{Department of Physical Chemistry, University of the Basque Country UPV/EHU, Apartado 644, E-48080 Bilbao, Spain}
\author{J. S. Pedernales}
\affiliation{Department of Physical Chemistry, University of the Basque Country UPV/EHU, Apartado 644, E-48080 Bilbao, Spain}
\author{I. L. Egusquiza}
\affiliation{Department of Theoretical physics and History of Science, University of the Basque Country UPV/EHU, Apartado 644, E-48080 Bilbao, Spain}
\author{G. Romero}
\affiliation{Departamento de F\'isica, Universidad de Santiago de Chile (USACH), Avenida Ecuador 3493, 917-0124, Santiago, Chile}
\author{L. Lamata}
\affiliation{Department of Physical Chemistry, University of the Basque Country UPV/EHU, Apartado 644, E-48080 Bilbao, Spain}
\author{D. Braak}
\affiliation{Experimental Physics VI, Center for Electronic Correlations and Magnetism, University of Augsburg, 86135 Augsburg, Germany}
\affiliation{Center for Correlated Matter, Zhejiang University, Hangzhou 310058, China}
\author{E. Solano}
\affiliation{Department of Physical Chemistry, University of the Basque Country UPV/EHU, Apartado 644, E-48080 Bilbao, Spain}
\affiliation{IKERBASQUE, Basque Foundation for Science, Maria Diaz de Haro 3, 48013 Bilbao, Spain}

\begin{abstract} 
Two-photon processes have so far been considered only as resulting from frequency-matched second-order expansions of light-matter interaction, with consequently small coupling strengths. However, a variety of novel physical phenomena arises when such coupling values become comparable with the system characteristic frequencies. Here, we propose a realistic implementation of two-photon quantum Rabi and Dicke models in trapped-ion technologies. In this case, effective two-phonon processes can be explored in all relevant parameter regimes. In particular, we show that an ion chain under bichromatic laser drivings exhibits a rich dynamics and highly counterintuitive spectral features, such as interaction-induced spectral collapse.
\end{abstract}

\date{\today}

\maketitle
\section{Introduction}
The quantum Rabi model describes the interaction of a two-level quantum system, a qubit, with a quantized single-mode bosonic field. Its semiclassical limit, where a classical field is considered, is known as the Rabi model~\cite{Rabi1937}. In the last few decades, the quantum Rabi model has been used in a regime where the rotating-wave approximation (RWA) holds, giving rise to the Jaynes-Cummings model~\cite{JaynesCummings} and describing a plethora of experiments, mostly related to cavity quantum electrodynamics. On the other hand, from a mathematical point of view, an analytical solution for the spectrum of the quantum Rabi model has been recently developed~\cite{Braak2011}. Such results have prompted a number of theoretical efforts aimed at applying similar techniques to generalizations of the quantum Rabi model, including anisotropic couplings~\cite{Xie2014, Cui2015, Zhong2014}, two-photon interactions~\cite{Travenec2012, Albert2011} and multi-qubit extensions~\cite{Braak2013}, as is the case of the Dicke model.

 In particular, the two-photon quantum Rabi model enjoys a spectrum with highly counterintuitive features~\cite{Ng1999, Emary2002}, which appear when the coupling strength becomes comparable with the bosonic mode frequency. In this sense, it is instructive to compare these features with the ultrastrong~\cite{Bourassa2009,Niemczyk2010,Diaz2010} and deep strong~\cite{Casanova2010} coupling regimes of the quantum Rabi model~\cite{Pedernales2015}. The two-photon Rabi model has been applied as an effective model to describe second-order processes in different physical setups, such as Rydberg atoms in microwave superconducting cavities \cite{Bertet2002} and quantum dots \cite{Stufler2008, delValle2010}. However, the small second-order coupling strengths restrict the observation of a richer dynamics.

In trapped-ion systems~\cite{leibfried2003,Haffner2008}, it is possible to control the coherent interaction between the vibrations of an ion crystal and internal electronic states, which form effective spin degrees of freedom. This quantum technology has emerged as one of the most promising platforms for the implementation of  quantum spin models, including few~\cite{Kim2009} or hundreds~\cite{Britton2012} of ions. Disparate complex quantum phenomena have been explored using trapped-ion setups, such as Ising spin frustration~\cite{Kim2010}, quantum phase transitions~\cite{Friedenauer2008,Islam2011} and the inhomogeneous Kibble-Zurek mechanism~\cite{delCampo2010}. Furthermore, second sidebands have been considered for laser cooling~\cite{Filho1994} and for generating nonclassical motional states~\cite{Meekhof1996, Filho1996, Gou1996}.

In this paper, we design a trapped-ion scheme in which the two-photon Rabi and two-photon Dicke models can be realistically implemented in all relevant regimes. We theoretically show that the dynamics of the proposed system is characterized by harmonic two-phonon oscillations or by spontaneous generation of excitations, depending on the effective coupling parameter. In particular, we consider cases where complete spectral collapse---namely, the fusion of discrete energy levels into a continuous band---can be observed.

\section{The model}
We consider a chain of $N$ qubits interacting with a single bosonic mode via two-photon interactions
\begin{equation}
\mathcal{H} = \omega a^\dagger a + \sum_n \frac{\omega_q^n}{2} \sigma_z^n + \frac{1}{N}\sum_n g_n \sigma_x^n \left( a^2 + {a^\dagger}^2 \right),
\label{2phdicke}
\end{equation}
where $\hbar=1$, $a$ and $a^\dagger$ are bosonic ladder operators; $\sigma_x^n$ and $\sigma_z^n$ are qubit Pauli operators; parameters $\omega$, $\omega_q^n$, and $g_n$, represent the mode frequency, the $n$-th qubit energy spacing and the relative coupling strength, respectively. We will explain below how to implement this model using current trapped-ion technology, considering in detail the case $N=1$ and discussing the scalability issues for $N > 1$.

We consider a setup where the qubit energy spacing, $\omega_{\rm int}$, represents an optical or hyperfine/Zeeman internal transition in a single trapped ion. The vibrational motion of the ion is described by bosonic modes $a,\ a^\dagger$, with trap frequency $\nu$. Turning on a bichromatic driving, with frequencies $\omega_r$ and $\omega_b$, an effective coupling between the internal and motional degrees of freedom is activated. In the interaction picture, the standard Hamiltonian~\cite{leibfried2003} describing this model reads
\begin{equation}
\mathcal{H}^I= \sum_{j=r,b} \frac{\Omega_j}{2} \left\{ e^{i\eta_j \left[ a(t) + a^\dagger(t) \right]  } e^{i \left(\omega_{\rm int} - \omega_j \right)t} e^{i \phi_j} \sigma_+ + \text{H.c.}
\right\},
\label{fundINT}
\end{equation}
where $a(t) = a\ e^{-i\nu t}$. Here, $\Omega_r$ and $\Omega_b$ are coupling parameters directly proportional to the driving laser amplitude, and $\phi_j$ represents the phase of each laser with respect to the atom dipole.
The Lamb-Dicke parameter $\eta_j = k^j_z \sqrt{\frac{\hbar}{2 m \nu}}$ is defined by the the projection $k^j_z$ of the $j$-th laser field wavevector in the $z$ direction and by the ion mass $m$. We consider the system to be in the Lamb-Dicke regime, $\eta^2 (2 \langle \hat{n} \rangle +1) \ll 1$, where $\hat{n} = a^\dagger a$ is the phonon number operator.

We set the frequencies of the bichromatic driving to be detuned from the second sidebands, 
$\omega_r = \omega_{\rm int} - 2\nu + \delta_r$, $\omega_b = \omega_{\rm int} +2 \nu + \delta_b$.
We choose  homogeneous Lamb-Dicke parameters $\eta_j= \eta$, phases $\phi_j=0$, and coupling strengths $\Omega_j = \Omega$ for both sideband excitations.  Expanding the exponential operator in Eq.~\eqref{fundINT} to the second order in $\eta$, and performing a RWA with $\delta_j, \Omega_j \ll \nu$, we can rewrite the interaction picture Hamiltonian
\begin{equation}
\mathcal{H}^I= -\frac{\eta^2\Omega}{4}\left[a^2\ e^{-i \delta_r t} + {a^\dagger}^2\ e^{-i\delta_b t} \right] \sigma_+ + \text{H.c.}
\label{rwaINT}
\end{equation}
The first-order correction to approximations made in deriving Eq.~\eqref{rwaINT} is given by $\frac{\Omega}{2}\ e^{\pm i2\nu t}\sigma_+ + \text{H.c.}$, which produce spurious excitations with negligible probability $P_e = \left( \frac{\Omega}{4\nu} \right)^2$. Further corrections are proportional to $\eta \Omega$ or $\eta^2$ and oscillate at frequency $\nu$, yielding $P_e = \left( \frac{\eta\Omega}{4\nu} \right)^2$. Hence, they are negligible in standard trapped-ion implementations. The explicit time dependence in Eq.~\eqref{rwaINT} can be removed by going to another interaction picture with $\mathcal{H}_0 = \frac{1}{4}\left(\delta_b - \delta_r \right)a^\dagger a + \frac{1}{4}\left( \delta_b + \delta_r\right) \sigma_z$, which we dub the simulation picture. Then, the system Hamiltonian resembles the two-phonon quantum Rabi Hamiltonian
\begin{equation}
\mathcal{H}_\text{eff} = \omega\ a^\dagger a + \frac{\omega_q}{2} \sigma_z - g\ \sigma_x \left( a^2 + {a^\dagger}^2 \right),
\label{implem}
\end{equation}
where the effective model parameters are linked to physical variables through $\omega = \frac{1}{4}\left( \delta_r - \delta_b \right)$, $\omega_q = -\frac{1}{2}\left( \delta_r + \delta_b \right)$, and $g=\frac{\eta^2 \Omega }{4}$. Remarkably, by tuning $\delta_r$ and $\delta_b$, the two-phonon quantum Rabi model of Eq.~\eqref{implem} can be implemented in all regimes. Moreover, the {\it N}-qubit two-phonon Dicke model of Eq.~\eqref{2phdicke} can be implemented using a chain of {\it N} ions by applying a similar method. In this case, the single bosonic mode is represented by a collective motional mode~\cite{James1997} (see Appendix~\ref{app_dicke}).

\begin{figure}[t]
\includegraphics[width=0.45\textwidth]{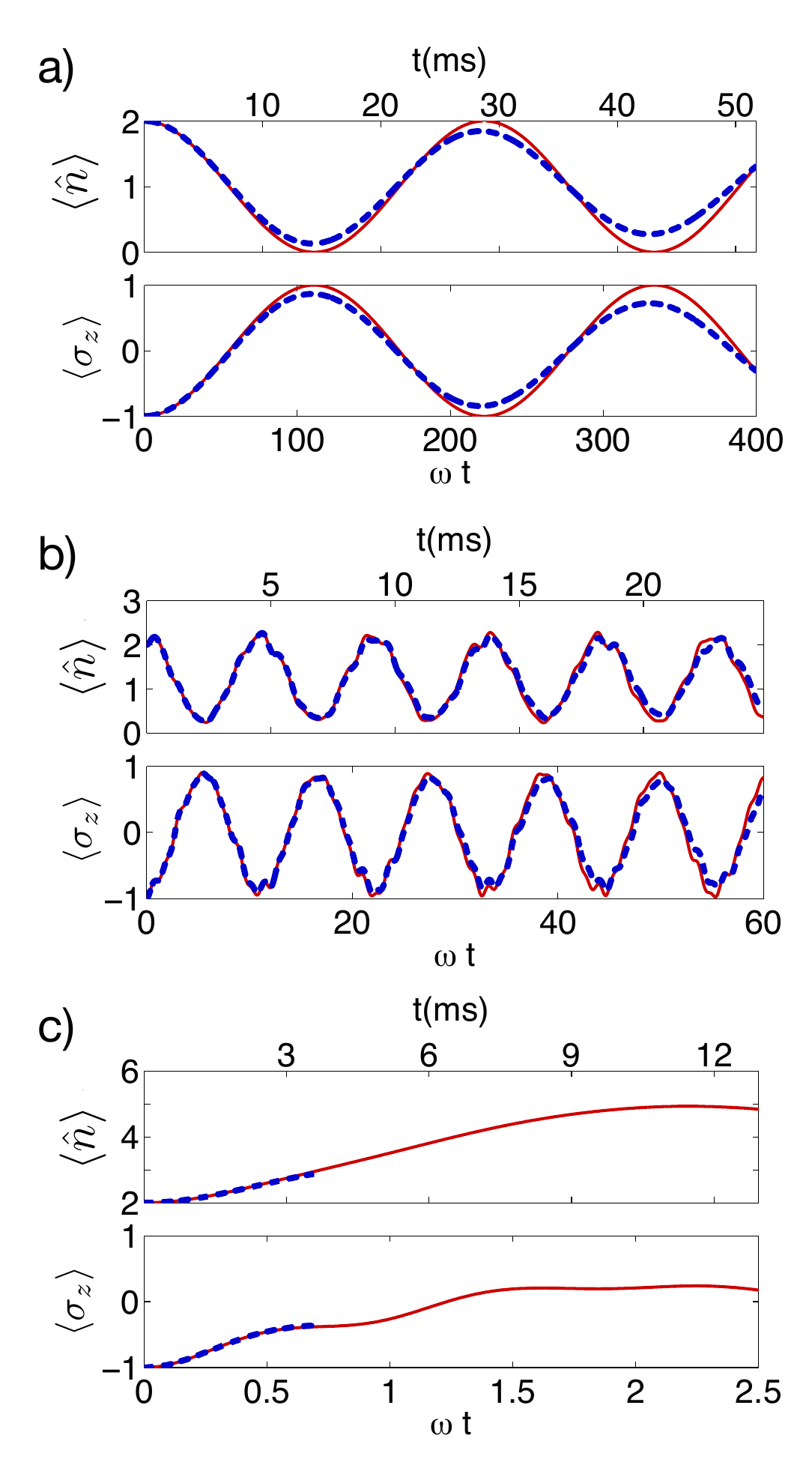}
\caption{(color online) Real-time dynamics for $N=1$, resonant qubit $\omega_q = 2\omega$, and effective couplings: (a) $g=0.01\omega$, (b) $g=0.2\omega$, and (c) $g=0.4\omega$. The initial state is given by $\ket{g, 2}$, i.e., the two-phonon Fock state and the qubit ground state. In all plots, the red solid line corresponds to numerical simulation of the exact Hamiltonian of Eq.~\eqref{2phdicke}, while the blue dashed line is obtained simulating the full model of Eq.~\eqref{fundINT}, including qubit decay $t_1 = 1$s, pure dephasing $t_2 = 30$ms and vibrational heating of one phonon per second. In each plot, the lower abscissa shows the time in units of $\omega$, while the upper one shows the evolution time of a realistic trapped-ion implementation. In panel (c), the full model simulation could not be performed for a longer time due to the fast growth of the Hilbert-space. }
\label{realtime}
\end{figure}

The validity of the approximations made in deriving Eq.~\eqref{implem} has been checked comparing the simulated two-photon quantum Rabi dynamics with numerical evaluation of the simulating trapped-ion model of Eq.~\eqref{fundINT}, as shown in Fig.~\ref{realtime}. Standard parameters and dissipation channels of current setups have been considered. In all plots of Fig.~\ref{realtime},  the vibrational frequency is $\nu/2\pi = 1$~MHz and the coupling coefficient is $\Omega/2\pi = 100$~KHz. The Lamb-Dicke parameter is $\eta=0.04$ for Fig.~\ref{realtime}a and Fig.~\ref{realtime}b, while $\eta=0.02$ for Fig.~\ref{realtime}c. Notice that larger coupling strengths imply a more favourable ratio between dynamics and dissipation rates. Hence, the implementation accuracy improves for large values of $g/\omega$, which correspond to the most interesting coupling regimes.   
 
\section{Real-time dynamics}
Depending on the ratio between the normalized coupling strength $g$ and the mode frequency $\omega$, the model of  Eq.~\eqref{2phdicke} exhibits qualitatively different behaviors. Two parameter regimes can be identified accordingly. 
For the sake of simplicity, we will consider the homogeneous coupling case $g_n=g$, $\omega_q^n = \omega_q$, for every $n$, and we will focus
 in the resonant or near-resonant case $\omega_q \approx 2\omega$. 

In accordance with the quantum Rabi model, we define the strong coupling (SC) regime by the condition $g/\omega \ll1$. Under this restriction, the RWA can be applied to the coupling terms, replacing each direct interaction $ g \sum_n \sigma_x^n \left( a^2 + {a^\dagger}^2 \right)$ with  $ g \sum_n \left( \sigma_+^n a^2 + \sigma_-^n{a^\dagger}^2 \right)$, where we defined the raising/lowering single-qubit operators $\sigma^n_\pm = \left(\sigma^n_x \pm i\sigma^n_y \right)/2$. 
When the RWA is valid, the system satisfies a continuous symmetry, identified by the operator $\zeta = a^\dagger a + 2\sum_n \sigma_+^n\sigma_-^n$, which makes the model superintegrable~\cite{Braak2011}.  In the SC regime, the interaction leads to two-photon excitation transfers between the bosonic field and the qubits, as shown in Fig.~\ref{realtime}a. 
Jaynes-Cummings-like collapses and revivals of population inversion are also expected to appear~\cite{Alsing1986, Joshi2000}.

As the ratio $g/\omega$ increases, the intuitive dynamics of the SC regime disappears and excitations are not conserved  (see Fig.~\ref{realtime}b and Fig.~\ref{realtime}c).
When the normalized coupling approaches the value $g\sim 0.1\omega$, the RWA cannot be performed, and the full quantum Rabi model must be taken into account.  We define the ultrastrong coupling (USC) regime as the parameter region for which $0.1 \lesssim g/\omega < 0.5$.
An analytical solution for the system eigenstates has  been derived in~\cite{Travenec2012}. However, this approach relies basically on a numerical instability related to the presence of dominant and minimal solutions of an associated three-term recurrence relation~\cite{Zhang2014} and gives no qualitative insight into the behavior of the spectrum close to the collapse point. While continued-fraction techniques are applicable in principle~\cite{Zhang2014}, only a few low-lying levels can be computed and the method fails again in approaching the critical coupling (see below). While the $G$-function derived in~\cite{Chen2012} allows for the desired understanding of the qualitative features of the collapse, its mathematical justification is still incomplete. On the other hand, direct numerical simulation becomes challenging close to collapse due to the large number of excitations involved. Especially the dynamics of the two-photon Dicke model is demanding for classical numerical techniques.

\begin{figure}[]
\includegraphics[width=0.45\textwidth]{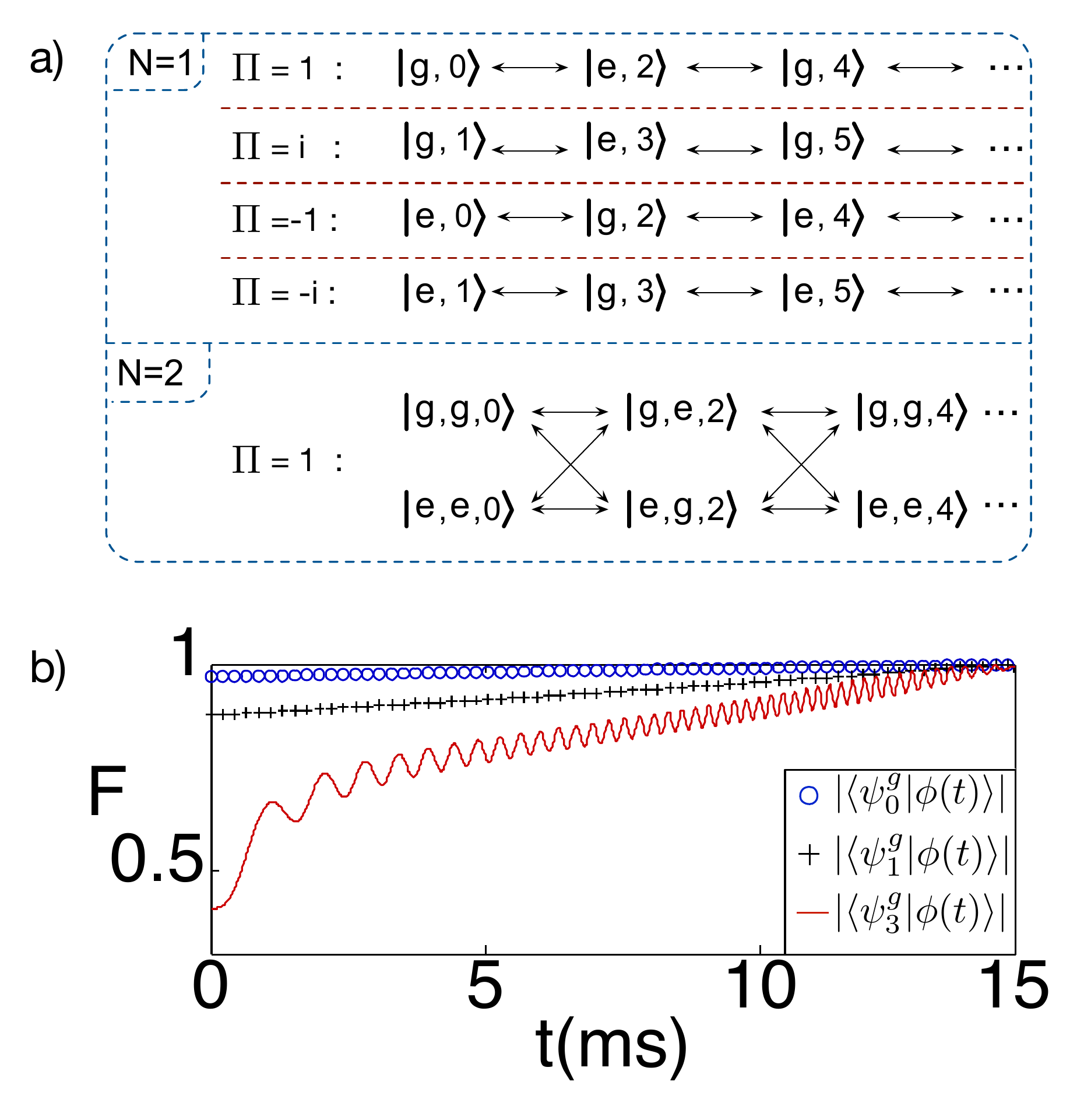}
\caption{(color online) (a) Generalized-parity chains for $N=1, 2$. For simplicity, for $N=2$, only one generalized-parity subspace is shown. (b) Quantum  state fidelity between the system state $\ket{\phi(t)}$ and the target eigenstates $\ket{\psi^g_n}$ during adiabatic evolution. The Hamiltonian at $t=0$ is given by Eq.~\eqref{2phdicke} with $N=1$, $\omega_q/\omega = 1.9$ and $g=0$. During the adiabatic process, the coupling strength is linearly increased until reaching the value $g/\omega = 0.49$. For the blue circles, the initial state is given by the ground state $\ket{\phi(t=0)} = \ket{\psi^{g=0}_0}$. For black crosses, $\ket{\phi(t=0)} = \ket{\psi^{g=0}_1}$, while for the red solid line, $\ket{\phi(t=0)} = \ket{\psi^{g=0}_4}$. The color code indicates  generalized parity as in Fig.~\ref{spectrum}a. Notice that, due to generalized parity conservation, the fourth excited eigenstate $\ket{\psi^{g=0}_4}$  of the decoupled Hamiltonian is transformed into the third one $\ket{\psi^{g}_3}$ of the full Hamiltonian.}
\label{paritychain}
\end{figure}

In the SC/USC transition, the continuous symmetry $\zeta$ breaks down to a $\mathbb{Z}_4$ discrete symmetry identified by the  operator 
\begin{equation}
\Pi = (-1)^N \bigotimes_{n=1}^N \sigma_z^n\ \exp\left( i \frac{\pi}{2} a^\dagger a \right).
\label{parityoperator}
\end{equation} 
We will call $\Pi$ the generalized parity operator, in analogy with the standard quantum Rabi model~\cite{Braak2011}.
Four invariant Hilbert subspaces are identified by the four eigenvalues $\lambda = \{ 1,-1, i, -i \}$ of $\Pi$.
Hence, for any coupling strength, the symmetry $\Pi$  restricts the dynamics to generalized-parity chains, shown in Fig.~\ref{paritychain}a, for $N=1,2$. 

When the normalized coupling $g$ approaches $g = \omega /2 $ (see Fig~\ref{spectrum}c), the dynamics is dominated by the interaction term and it is characterized by photon production. Finally, when  $g > \omega/2$, the Hamiltonian is not bounded from below. However, it still provides a well defined dynamics when applied for a limited time, like usual displacement or squeezing operators.

\section{The spectrum}
\begin{figure}[h]
\includegraphics[width=0.45\textwidth]{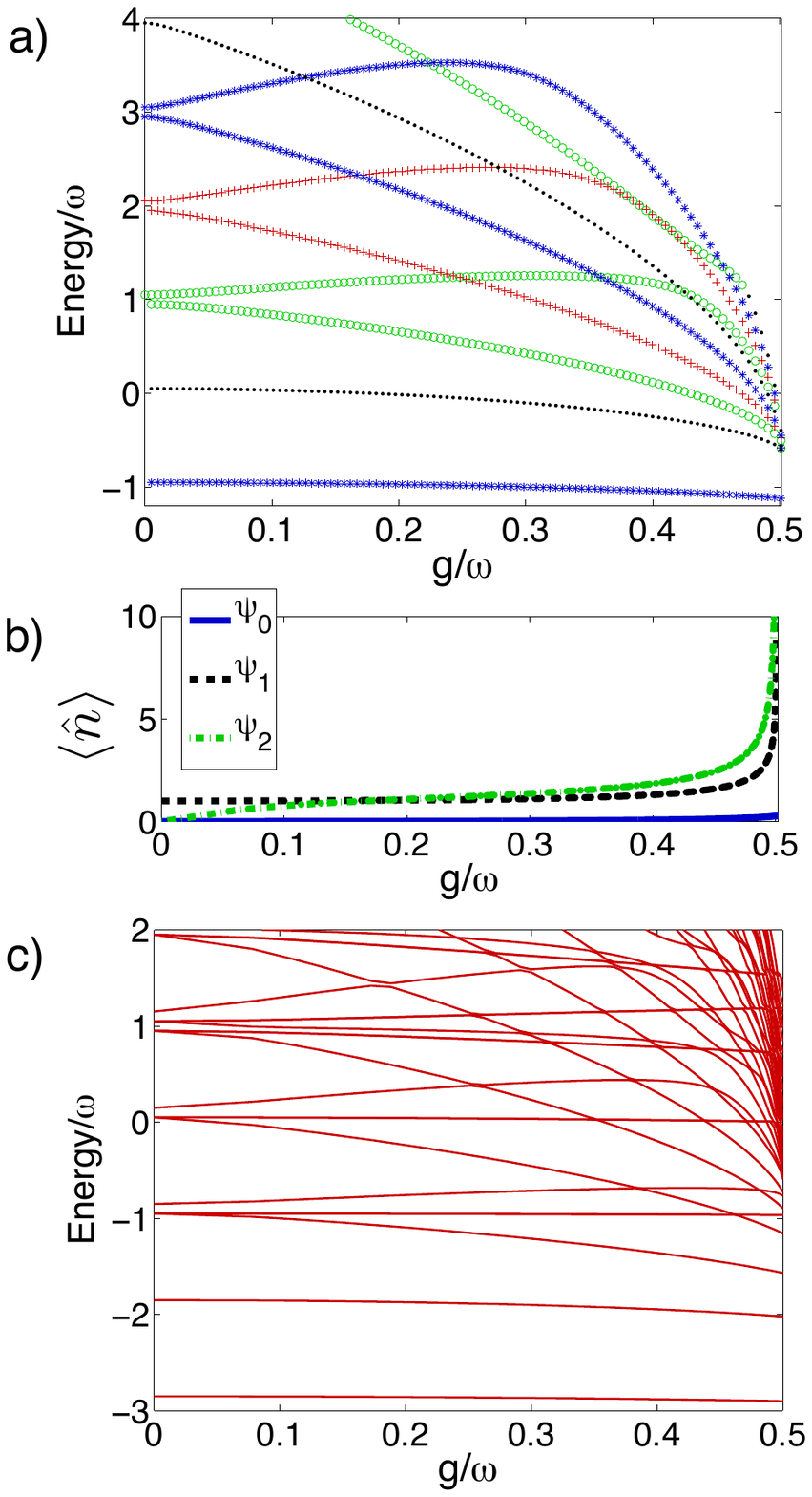}
\caption{(color online)  Spectral properties of the Hamiltonian Eq.~\eqref{2phdicke}, in units of $\omega$, for $\omega_q = 1.9$, as a function of the coupling strength $g$. For $g>0.5$, the spectrum is unbounded from below. (a) Spectrum for $N=1$. Different markers identify the generalized parity of each eigenstate: green circles for $p=1$, red crosses for $p=i$, blue stars for $p=-1$, and black dots for $p=-i$. (b) Average photon number for the ground and first two excited states, for $N=1$. (c) Spectrum for $N=3$. For clarity, the generalized parity of the eigenstates is not shown. }
\label{spectrum}
\end{figure}

  The eigenspectrum of the Hamiltonian in Eq.~\eqref{2phdicke} is shown in Figs.~\ref{spectrum}a and \ref{spectrum}c for $N=1$ and $N=3$, respectively. Different markers are used to identify the generalized parity $\Pi$ of each Hamiltonian eigenvector, see Eq.~\eqref{parityoperator}. In the SC regime, the spectrum is characterized by the linear dependence of the energy splittings, observed for small values of $g$.  On the contrary, in the USC regime the spectrum is characterized by level crossings known as Juddian points, allowing for closed-form isolated solutions~\cite{Emary2002} in the single-qubit case.

The most interesting spectral features appear when the normalized coupling $g$ approaches the value $\omega/2$. In this case, the energy spacing between the system eigenenergies asymptotically vanishes and the average photon number  for the first excited eigenstates diverges (see Fig.~\ref{spectrum}b). When $g = \omega/2$, the discrete spectrum collapses into a continuous band, and its eigenfunctions are not normalizable (see Appendix~\ref{app_math}). Beyond that value, the Hamiltonian is unbounded from below~\cite{Ng1999, Emary2002}. This can be shown by rewriting the bosonic components of Hamiltonian of Eq.~\eqref{2phdicke} in terms of the effective position and momentum operators of a particle of mass $m$, defined as $\hat{x} = \sqrt{\frac{1}{2m\omega}}\left( a + a^\dagger \right)$ and $\hat{p} = i\sqrt{\frac{m\omega}{2}}\left( a - a^\dagger \right)$. Therefore, we obtain
\begin{eqnarray}
\mathcal{H} &=& \frac{m\omega}{2}\left[ (\omega - 2g\ \hat{S}_x )\frac{\hat{p}^2}{m^2 \omega^2} + (\omega + 2g\ \hat{S}_x ) \hat{x}^2 \right]  \nonumber \\
&+&  \frac{\omega_q}{2}  \sum_n \sigma_z^n,
\label{effH}
\end{eqnarray}
where $\hat{S}_x = \frac{1}{N}\sum_n \sigma_x^n$.  Notice that  $\hat{S}_x$ can take values included in the interval $\langle S_x\rangle \in [-1,1]$. Hence, the parameter $(\omega + 2g)$ establishes the shape of the effective potential. For $g < \omega/2$, the particle experiences an always  positive quadratic potential. For $g = \omega/2$, there are qubit states which turn the potential flat and the spectrum collapses, like for a free particle (see Appendix~\ref{app_math}). Finally, when $g>\omega/2$, the effective quadratic potential can be positive, for $\langle\hat{S}_x\rangle < -\omega/2g$, or negative, for $\langle\hat{S}_x\rangle >  \omega/2g$. Therefore, the Hamiltonian~\eqref{effH} has neither an upper nor a lower bound.

\section{Measurement technique}
A key experimental signature of the spectral collapse (see Fig.~\ref{spectrum}a) can be obtained by measuring the system eigenenergies~\cite{Senko2014} when  $g$ approaches $0.5 \omega$. Such a measurement could be done via the quantum phase estimation algorithm~\cite{Abrams1999}. A more straightforward method consists of directly generating the system eigenstates~\cite{Felicetti2014} by means of the adiabatic protocol shown in Fig.~\ref{paritychain}b. When $g=0$, the eigenstates $\ket{\psi^{g=0}_n}$ of  Hamiltonian in Eq.~\eqref{2phdicke} have an analytical form and can be easily generated~\cite{Jurcevic}. Then, adiabatically increasing $g$, the eigenstates $\ket{\psi^{g}_n}$ of the full model can be produced. Notice that generalized-parity conservation protects the adiabatic switching at level crossings (see Fig.~\ref{paritychain}b).

Once a given eigenstate has been prepared, its energy can be inferred by measuring the expected value of the Hamiltonian in Eq.~\eqref{2phdicke}. We consider separately the measurement of each Hamiltonian term. The measurement of $\sigma^n_z$ is  standard in trapped-ion setups and is done with fluorescence techniques~\cite{leibfried2003}. The measurement of the phonon number expectation value was already proposed in Ref.~\cite{Bastin2006}. Notice that operators $\sigma^n_z$ and $a^\dag a$ commute with all transformations performed in the derivation of the model. The expectation value of the interaction term $g\sigma^n_x \left( a^2 + {a^\dag}^2\right)$ can be mapped into the value of the first time derivative of $\langle \sigma^n_z \rangle$ at measurement time $t=0$, with the system evolving under $\mathcal{H}_m=\omega a^\dag a + \frac{\omega_q}{2} \sigma^n_z - g \sigma^n_y \left( a^2 + {a^\dag}^2\right)$. This Hamiltonian is composed of a part $A=\omega a^\dag a + \frac{\omega_q}{2} \sigma^n_z$ which commutes with $\sigma^n_z$, $[A,\sigma^n_z]=0$, and a part $B=-g \sigma^n_y \left(a^2 + {a^\dag}^2\right)$ which anti-commutes with $\sigma^n_z$, $\{B,\sigma^n_z \}=0$, yielding
$\langle e^{i(A+B)t} \sigma^n_z e^{-i(A+B)t} \rangle = \langle e^{i(A+B)t} e^{-i(A-B)t}\sigma^n_z  \rangle$.
The time derivative of this expression at $t=0$ is given by
$\langle [i (A+B) - i (A - B)] \sigma^n_z \rangle = 2 i \langle B \sigma^n_z \rangle$,
which is proportional to the expectation value of the interaction term of Hamiltonian in Eq.~(\ref{implem}),
$\partial_t \langle e^{i\mathcal{H}_m t} \sigma^n_z e^{-i \mathcal{H}_m t} \rangle|_{t=0} =  2 \langle g \sigma^n_x \left(a^2 + {a^\dag}^2\right) \rangle.$
The evolution under Hamiltonian $\mathcal{H}_m$ in the simulation picture is implemented  in the same way as the Hamiltonian in Eq.~(\ref{implem}), but selecting the laser phases $\phi_j$ to be $\frac{\pi}{2}$.
Moreover, expectation values for the generalized-parity operator $\Pi$ of Eq.~\eqref{parityoperator} can be extracted following the techniques described in Appendix~\ref{app_meas}.

\section{Discussion}
We have introduced a trapped-ion scheme which allows one to experimentally investigate two-photon interactions in unexplored regimes of light-matter coupling, replacing photons in the model by trapped-ion phonons. It provides a feasible method to observe an interaction-induced spectral collapse in a two-phonon quantum Rabi model, approaching recent mathematical and physical results with current quantum technologies. Furthermore, the proposed scheme provides a scalable quantum simulator of a complex quantum system, which is difficult to approach with classical numerical simulations even for low number of qubits, due to the large number of phonons involved in the dynamics.
\begin{acknowledgments}
We acknowledge support from Basque government IT472- 281 10, Spanish MINECO FIS2012-36673-C03-02, Ram\'on y 282 Cajal Grant No. RYC-2012-11391, UPV/EHU PhD grant, 283 UPV/EHU UFI 11/55, UPV/EHU Project No. EHUA14/04, 284 Chilean FONDECYT 1150653, and the PROMISCE and 285 SCALEQIT European Union projects and partial support from 286 the National Natural Science Foundation of China, Grant No. 287 11474250, and German TRR80.
\end{acknowledgments}

\appendix

\section{Implementation of two-photon Dicke model with collective motion of $N$ trapped ions}
\label{app_dicke}
In the main text, we showed how a two-photon Rabi model can be implemented using the vibrational degree of freedom of a single ion, coupled to one of its internal electronic transitions by means of laser-induced interactions. Here, we show how the N-qubit two-photon Dicke model
\begin{equation}
\mathcal{H} = \omega a^\dagger a + \sum_n \frac{\omega_q^n}{2} \sigma_z^n + \frac{1}{N}\sum_n g_n \sigma_x^n \left( a^2 + {a^\dagger}^2 \right),
\label{sup_2phdicke}
\end{equation}
can be implemented in a chain of $N$ ions, generalizing such a method. The $N$ qubits are represented by an internal electronic transition of each ion, while the bosonic mode is given by a collective motional mode of the ion chain. The two-phonon interactions are induced by a bichromatic laser driving with the same frequency-matching conditions used for the single-qubit case. The drivings can be implemented by shining two longitudinal lasers coupled to the whole chain, or by addressing the ions individually with transversal beams. The former solution is much less demanding, but it may introduce inhomogeneities in the coupling for very large ion chains; the latter allows complete control over individual coupling strengths.

In order to guarantee that the model of Eq.~\eqref{sup_2phdicke} is faithfully implemented, the bichromatic driving must not excite unwanted motional modes. In our proposal, the frequency of the red/blue drivings $\omega_{r/b}$ satisfy the relation $|\omega_{r/b} - \omega_{int} | = 2\nu + \delta_{r/b}$, where $\delta_{r/b}$ are small detunings that can be neglected for the present discussion. We recall that $\nu$ is the bosonic mode frequency and $\omega_{int}$ the qubit energy spacing. To be definite, we take the motion of the center of mass of the ion chain as the relevant bosonic mode. Then, the closest collective motional mode is the breathing mode~\cite{James1997}, with frequency $\nu_2 = \sqrt{3}\nu$. An undesired interaction between the internal electronic transitions and the breathing mode could appear if $|\omega_{r/b} - \omega_{int} |$ is close to $\nu_2$ or $2\nu_2$, corresponding to the first and second sidebands, respectively. In our case, the drivings are detuned by $\Delta_1 = |\omega_{r/b} - \omega_{int} | - \nu_2 \approx 0.27 \nu$  from the first and $\Delta_2 = |\omega_{r/b} - \omega_{int} | - 2\nu_2 \approx 1.46 \nu$ from the second sideband. Given that the frequency $\nu$ is much larger than the coupling strength $\Omega$, such detunings make those unwanted processes safely negligible.

\section{Properties of the wavefunctions below and above the collapse point}
\label{app_math}
The presence of the collapse point at $g=\omega/2$ can be inferred rigorously by studying  the asymptotic behavior of the formal solutions to the time-independent Schr\"odinger equation $\mathcal{H}\psi=E\psi$. We consider now the simplest case $N=1$. Using the representation of the model in the Bargmann space $\mathcal{B}$ of analytic functions \cite{bargmann1961}, the Schr\"odinger equation for $\psi(z)$ in the invariant subspace with generalized-parity eigenvalue $\Pi=+1$ reads
\begin{equation}
g\psi''(z)+\omega z\psi'(z)+gz^2\psi(z) 
+ \frac{\omega_q}{2}\psi(iz)=E\psi(z),
\label{schroed-nonloc}
\end{equation}
where the prime denotes differentiation with respect to the complex variable $z$. This nonlocal linear differential equation of the second order, connecting the values of $\psi$ at the points $z$ and $iz$, may be transformed to a local equation of the fourth order,
\begin{widetext}
\begin{equation}
\psi^{(4)}(z)+[(2-\bar{\omega}^2)z^2+2\bar{\omega}]\psi''(z)
+[4+2\bar{\omega}\bar{E}-\bar{\omega}^2]z\psi'(z)+
[z^4-2\bar{\omega}z^2+2-\bar{E}^2+\Delta^2]\psi(z)=0,
\label{schroed-loc}
\end{equation}
\end{widetext}
where we have used the abbreviations $\bar{\omega}=\omega/g$, $\Delta=\omega_q/(2g)$,
$\bar{E}=E/g$. Equation ~\eqref{schroed-loc} has no singular points in the complex plane except at $z=\infty$, where it exhibits an unramified irregular singular point of s-rank three \cite{slavyanov2000}. That means that the so-called {\it normal solutions} have the asymptotic expansion
\begin{equation}
\psi(z)=e^{\frac{\gamma}{2}z^2+\alpha z}z^\rho(c_0+c_1z^{-1}+c_2z^{-2}+\ldots),
 \label{asym}
\end{equation}
for $z\rightarrow\infty$. 
Functions of this type are only normalizable (and belong therefore to $\mathcal{B}$) if the complex parameter $\gamma$, a characteristic exponent of the second kind, satisfies
$|\gamma|<1$. In our case, the possible $\gamma$'s are the solutions of the biquadratic 
equation
\begin{equation}
x^4+x^2(2-\bar{\omega}^2)+1=0.
\label{biquad}
\end{equation}
It follows
\begin{equation}
\gamma_{1,2}=\frac{\bar{\omega}}{2}\pm\sqrt{\frac{\bar{\omega}^2}{4}-1}, \quad
\gamma_{3,4}=-\frac{\bar{\omega}}{2}\pm\sqrt{\frac{\bar{\omega}^2}{4}-1}.
\label{sols}
\end{equation}
For $\bar{\omega}/2 >1$, all solutions are real. For $|\gamma_1|=|\gamma_4|>1$, we have $|\gamma_2|=|\gamma_3|<1$. In this case, there exist normalizable solutions if $\gamma_2$ or $\gamma_3$ appears in Eq.~\eqref{asym}. The condition for absence of the other characteristic exponents $\gamma_{1,4}$ in the formal solution of Eq.~\eqref{schroed-loc} is the spectral condition determining the parameter $E$ in the eigenvalue problem $\mathcal{H}\psi=E\psi$. It follows that for $g<\omega/2$, a discrete series of normalizable solutions to Eq.~\eqref{schroed-nonloc} may be found and the spectrum is therefore a pure point spectrum.

On the other hand, for $\bar{\omega}/2 <1$, all $\gamma_j$ are located on the unit circle with $\gamma_1=\gamma_2^\ast, \gamma_3=\gamma_4^\ast$. Because, then, no normalizable solutions of Eq.~\eqref{schroed-loc} exist, the spectrum of the (probably self-adjoint) operator $\mathcal{H}$ must be continuous for $g>\omega/2$, i.e. above the collapse point. The exponents $\gamma_1$ and $\gamma_2$ ($\gamma_3$ and $\gamma_4$) join at 1 (-1) for $g=\omega/2$. The exponent $\gamma=1$ belongs to the Bargmann representation of plane waves. Indeed, the plane wave states $\phi_q(x)=(2\pi)^{-1/2}\exp(iqx)$ in the rigged extension of $L^2(\mathbb{R})$ \cite{gelfand1964}, satisfying the othogonality relation $\langle\phi_q|\phi_{q'}\rangle=\delta(q-q')$, are mapped by the isomorphism $\mathcal{I}$ between $L^2(\mathbb{R})$ and $\mathcal{B}$ onto the functions
\begin{equation}
\mathcal{I}[\phi_q](z)=\pi^{-1/4}e^{-\frac{1}{2}q^2+\frac{1}{2}z^2+i\sqrt{2}qz}, 
\end{equation}   
they correspond therefore to $\gamma=1$. It is yet unknown whether at the collapse point $g=\omega/2$, the generalized eigenfunctions of $\mathcal{H}$ have plane wave characteristics for $\omega_q\neq 0$ or which properties of these functions appear above this point, where the spectrum is unbounded from below.

\section{Generalized-parity measurement}
\label{app_meas}
The generalized-parity operator, defined as $\Pi=(-1)^N\bigotimes_{n=1}^N \sigma_z^n {\rm exp}\{i\frac{\pi}{2} n \}$, with $n=a^\dag a$, is a non-Hermitian operator that can be explicitly written as the sum of its real and imaginary parts,
\begin{eqnarray}
\Pi &=& (-1)^N \bigotimes_{n=1}^N \sigma_z^n \cos(\frac{\pi}{2} a^\dag a) \\
&+& i (-1)^N \bigotimes_{n=1}^N \sigma_z^n \sin ({\frac{\pi}{2}a^\dag a}). \nonumber
\end{eqnarray}
For simplicity, we will focus on the $N=1$ case, but the procedure is straightforwardly extendible to any $N$. We will show how to measure the expectation value of operators of the form 
\begin{equation}
\exp \{ \pm i n \  \sigma_i \  \phi \} \sigma_j,
\label{accesible operator}
\end{equation} 
where $\sigma_{i,j}$ are a pair of anti-commuting Pauli matrices, $\{ \sigma_i, \sigma_j \}=0$, and $\phi$ is a continuous real parameter. One can then reconstruct the real and imaginary parts of the generalized-parity operator, as a composition of observables in Eq.~(\ref{accesible operator}) for different signs and values of $i, j$,
\begin{eqnarray}
\Re (\Pi)&=& -\frac{1}{2} \{ \exp ( i n \sigma_{x} \frac{\pi}{2} ) \sigma_z + \exp ( - i n \sigma_{x} \frac{\pi}{2} ) \sigma_z \}, \\
\Im (\Pi) &=& \frac{1}{2} \{ \exp ( i n \sigma_{x} \frac{\pi}{2} ) \sigma_{y} - \exp ( - i n \sigma_{x} \frac{\pi}{2} ) \sigma_{y} \}.
\end{eqnarray}

The strategy to retrieve the expectation value of observables in Eq.~(\ref{accesible operator}) will be based on the following property of anti-commuting matrices $A$ and $B$: $e^{A}Be^{-A}=e^{2A}B=Be^{-2A}$. Based on this, the expectation value of the observables in Eq.~(\ref{accesible operator}) can be mapped onto the expectation value of $\sigma_j$ when the system has previously evolved under Hamiltonian $H=\pm n \sigma_i $ for a time $t^*=\phi/2$,
\begin{equation}
\langle \psi | \exp \{ \pm i n \  \sigma_i \  \phi \} \sigma_j | \psi \rangle = \langle \psi (t^*) | \sigma_j | \psi (t^*) \rangle,
\end{equation}
where $ | \psi(t) \rangle = e^{-i n \sigma_i t} | \psi \rangle$. The expectation value of any Pauli matrix is accesible in trapped-ion setups, $\sigma_z$ by fluorescence techniques and $\sigma_{x,y}$ by applying rotations prior to the measurement of $\sigma_z$. The point then is how to generate the dynamics of Hamiltonian $H=\pm n \sigma_i $. For that, we propose to implement a highly detuned simultaneous red and blue sideband interaction,
\begin{equation}
H= \frac{ \Omega_0 \eta}{2} (a + a^\dag) \sigma^+ e^{i \delta t} e^{i\varphi}+ {\rm H.c.},
\end{equation}
where $\varphi$ is the phase of the laser with respect to the dipole moment of the ion. This Hamiltonian can be effectively approximated to the second-order Hamiltonian,
\begin{equation}
H_{\rm eff}= \frac{1}{\delta}\Big( \frac{ \Omega_0 \eta}{2} \Big)^2 (2n + 1) \sigma_z e^{i \varphi} ,
\label{Effective Hamiltonian}
\end{equation}
when $\delta \gg \eta \Omega_0/2$. The laser phase will allow us to select the sign of the Hamiltonian. Of course, one would need to be careful and maintain $\delta$ in a regime where $\delta \ll \nu$, $\nu$ being the trapping frequency, to guarantee that higher-order resonances are not excited. Finally, in order to get rid of the undesired extra term $\sigma_z$ in Hamiltonian Eq.~\eqref{Effective Hamiltonian}, one needs to implement one more evolution under the Hamiltonian $H= - (1/2) \Omega_0 \eta \sigma_z$. This evolution can be generated by means of a highly detuned carrier transition. So far, we have given a protocol to generate the Hamiltonian $H=\pm n \sigma_z$. In order to generate Hamiltonians $H=\pm n \sigma_{y}$, one would need to modify the evolution with two local qubit rotations, 
\begin{equation}
e^{ \pm i  n \sigma_{y}t}= e^{i \sigma_{x} \pi/4} e^{ \pm i n \sigma_z t} e^{-i \sigma_{x} \pi/4}.
 \end{equation}
 Similarly, for Hamiltonian $H=\pm n \sigma_x$ one would need to perform rotations around $\sigma_y$.

\end{document}